\documentclass[12 pt]{amsart}
\usepackage{graphicx}

\usepackage{amsthm,amsmath,latexsym,amssymb,amsfonts,amscd,hyperref}
\usepackage{color,tikz,tikz-cd}
\usepackage[all]{xy}

\theoremstyle{definition}

\usepackage{epigraph}

\theoremstyle{remark}

\numberwithin{equation}{section}

\begin{document}
\phantom{a}
\title[A note on quantum Hamiltonian reduction and anomalies]{A note on quantum Hamiltonian reduction and anomalies}

\author{Boris M. Elfimov}
\address{Physics Faculty, Tomsk State University, Lenin ave. 36, Tomsk 634050, Russia}
\email{e1fimov@mail.ru}
\author{ Alexey A. Sharapov }
\address{Physics Faculty, Tomsk State University, Lenin ave. 36, Tomsk 634050, Russia}
\email{sharapov@phys.tsu.ru}





\begin{abstract}
Quantization of field-theoretic models with gauge symmetries is often obstructed by quantum anomalies. It is commonly believed that the origin of these anomalies lies in the infinite number of degrees of freedom, which requires completing the model within an appropriate regularization scheme. This paper provides an explicit example of a finite-dimensional Hamiltonian system with first-class constraints whose quantization exhibits anomalies. These anomalies arise from the nontrivial topology of the reduced phase space.

\end{abstract}

\maketitle

\epigraph{ \it To the cherished memory of Hans-Christian Herbig, gone too soon}

\section{Introduction}
The term {\it quantum anomaly} broadly refers to the violation of classical properties upon quantization. In quantum field theory, it typically implies the breaking of global or local symmetries present in the classical model. Quantum anomalies -- or, more precisely, the conditions required for their cancellation -- play a crucial role in understanding fundamental interactions. For instance, the requirement of conformal anomaly cancellation in string theory imposes stringent constraints on the spacetime dimension and other physical parameters. Similarly, the absence of a chiral anomaly in the Standard Model   entails  exact equality between the number of quark and lepton generations.

Despite their diversity, all anomalies share a common origin related to the infinite dimensionality of field space. The infinite number of degrees of freedom necessitates regularization of a field-theoretical model, which may violate its classical symmetries. If the broken symmetry is not restored when the regulator is removed, an anomaly arises. In this work,  by contrast, we focus on quantum anomalies within finite-dimensional gauge systems.

There are numerous approaches to studying quantum anomalies in gauge theories. The most general and consistent framework is provided by the Hamiltonian formalism and the canonical quantization method. From this perspective, all gauge theories are Hamiltonian systems with first-class constraints. The quantization of such systems was initiated by Dirac and further developed into its modern form by Batalin,  Fradkin, and Vilovisky (see, for example, \cite{Dirac, BF, HT}). Within the BFV formalism, gauge anomalies manifest themselves as violations of the so-called master equations.

Many claims regarding quantum anomalies in the physics literature unfortunately lack mathematical rigor.
 In the context of field theory, the question of whether anomalies depend on the choice of regularization scheme remains open. Furthermore, studies of finite-dimensional Hamiltonian systems with constraints typically focus on local properties. The well-known thesis of Dirac's quantization \cite{Dirac}  -- ``$m$ functionally independent first-class constraints reduce the dimension of the physical phase space by $2m$'' -- is generally invalid when the global geometry of the phase space is taken into account. Simple counterexamples to this claim are discussed in the next section. Although the absence of quantum anomalies in finite-dimensional gauge systems was proved in \cite{BF}, this result also holds only locally.

This note aims to highlight global aspects of gauge theory quantization. General cohomological obstructions to the solutions of the quantum master equations have been studied in \cite{BHW, LS, LMS}. In this work, we construct an explicit example of a constrained Hamiltonian system that exhibits unavoidable quantum anomalies in the algebra of physical observables. The construction relies on general results from the theory of integrable Hamiltonian systems.

\section{Classical Reduction}

Let us consider a Hamiltonian system with $n$ degrees of freedom and $m$ first-class constraints $T_a\approx 0$.
In general, the phase space of the system is a $2n$-dimensional smooth manifold $M$ equipped with a non-degenerate Poisson bracket.
According to Dirac \cite{Dirac}, the first-class type of the Hamiltonian constraints means that their pairwise Poisson brackets vanish on the common constraint surface $\Sigma$:
\begin{equation}\label{inv}
\{T_a,T_b\}\approx 0\,.
\end{equation}
We will assume that the constraints satisfy the regularity condition
\begin{equation}\label{reg}
\left. dT_1\wedge\ldots \wedge dT_m \right|_\Sigma\neq 0\,.
\end{equation}
This condition guarantees that the constraint surface is a smooth coisotropic submanifold $\Sigma\subset M$ of dimension $2n-m$, and that the left-hand side of equation (\ref{inv}) is given by a linear combination of constraints $U^c_{ab}T_c$ with some smooth functions $U^c_{ab}$. The involution condition (\ref{inv}) means that the Hamiltonian vector fields $X_a=\{T_a, -\}$ are tangent to the constraint surface $\Sigma$, defining a smooth integrable distribution of rank $m$. The integral leaves of this distribution are called {\it gauge orbits}, and the resulting foliation   $\mathcal{F}$ on $\Sigma$ is  called the {\it gauge foliation}. In the case where the gauge foliation is a smooth fibration, the space of gauge orbits $\Sigma/\mathcal{F}$ is a smooth manifold (the base of the fibration), called the {\it reduced phase space} of the Hamiltonian system and denoted by $M/\!/{T_a}$. This terminology extends also to the general case where the set $\Sigma/\mathcal{F}$ is not a smooth Hausdorff manifold. In this case, “smooth functions” on the reduced phase space $N=M/\!/{T_a}$ are identified with smooth functions on $\Sigma$ that are constant along the gauge orbits:
\begin{equation}
C^\infty(N)=\big\{f\in C^{\infty}(\Sigma)\;|\; X_af=0\big\}\,.
\end{equation}
When the regularity condition (\ref{reg}) holds, the commutative algebra of functions $C^\infty(\Sigma)$ is isomorphic to the quotient algebra $C^\infty(M)/J$ by the ideal $J$ generated by the functions $T_a$. The involution condition (\ref{inv}) is equivalent to the closure of the ideal under Poisson brackets: $\{J,J\}\subset J$. This allows one to define the Poisson algebra of smooth functions on the reduced phase space as the quotient algebra
$C^{\infty}(M)^J/J$, where
\begin{equation}\label{CMJ}
C^{\infty}(M)^J=\big\{f\in C^\infty(M)\;|\; \{f,J\}\subset J\;\big\}
\end{equation}
is the stabilizer of $J$ in $C^\infty(M)$. With this definition, the original Poisson bracket on $M$ induces a Poisson structure on the algebra $C^\infty(N)$ even when the reduced phase space $N$ is not a smooth Hausdorff manifold.
Elements of the algebra $C^\infty(N)$ are called {\it physical observables} of the Hamiltonian system with constraints.

As an example illustrating both smooth and non-smooth Hamiltonian reduction by first-class constraints, consider the harmonic oscillator with two degrees of freedom. The Hamiltonian of the system has the form
\begin{equation}\label{H0}
H_0=H_1+H_2=\frac{\omega_1}2(p_1^2+x_1^2)+\frac{\omega_2}2(p_2^2+x_2^2)\,,
\end{equation}
where $\omega_{1}$ and $\omega_2$ are the frequencies of the oscillator, and $(p_1,x_1)$ and $(p_2,x_2)$ are canonically conjugate pairs of variables. We interpret the equation of the isoenergetic surface $H_0=E>0$ as a first-class constraint $H_0-E\approx 0$ in the 4-dimensional phase space $\mathbb{R}^4$. Topologically, the constraint surface $\Sigma$ is a 3-dimensional sphere (ellipsoid). The Hamiltonian flow $X_{H_0}=\{H_0,-\}$ does not vanish anywhere on $\Sigma$, foliating the constraint surface into one-dimensional gauge orbits. The structure of the resulting foliation is fully determined by whether the frequencies are commensurate or not: when $\omega_1/\omega_2\in \mathbb{Q}$ all trajectories are closed and diffeomorphic to $S^1$, otherwise all trajectories are diffeomorphic to $\mathbb{R}$. Additional information about the foliation structure can be obtained by noting that the energies of both oscillators, i.e., the functions $H_1$ and $H_2$, are first integrals of motion of the system (\ref{H0}). Clearly, the equations
\begin{equation}\label{H12}
H_1=E_1\,, \qquad H_2=E_2
\end{equation}
for various constants $E_1$ and $E_2$, related by $E_1+E_2=E$, foliate the isoenergetic surface of the oscillator into two-dimensional Liouville tori. These tori are all either resonant or non-resonant depending on whether the rotation number $\omega_1/\omega_2$ is rational or irrational. For the isotropic oscillator, for example, we immediately obtain the Hopf fibration $p: S^3\rightarrow S^2 $ with fiber $S^1$, so that the reduced phase space $\mathbb{R}^4/\!/\{H_0-E\}$ indeed turns out to be a smooth manifold diffeomorphic to $S^2$. In canonical action-angle variables $(s_1, \varphi_1 ; s_2,\varphi_2)$,
\begin{equation}\label{s}
s_i=\frac12(p^2_i+x^2_i)\,,\qquad \{s_i, \varphi_j\}=\delta_{ij}\,,\qquad \{\varphi_i,\varphi_j\}=0\,,
\end{equation}
each physical observable (smooth function on $S^2$) is represented by some function $f(s_1, \varphi_1-\varphi_2)$ of two variables, periodic in the second argument\footnote{Introducing spherical coordinates $(\theta, \phi)$ on $S^2$, one can set $s_1=E(1-\cos\theta)/2$ and $\phi=\varphi_1-\varphi_2$.}.

At the same time, for incommensurate frequencies, the topological closure of each integral trajectory (gauge orbit) yields a certain Liouville torus. As a result, the reduced phase space $\mathbb{R}^4/\!/\{H_0-E\}$ is not Hausdorff, and every function invariant under the Hamiltonian flow $X_{H_0}$ must be constant on each Liouville torus. The Fomenko graph \cite{Fom}, describing the evolution and bifurcation of Liouville tori at fixed $H_0$ and varying integral $H_1$, is a segment $0\leq E_1\leq E$. Each interior point of this segment corresponds to a regular torus; boundary points correspond to degenerate tori (circles) when one of the two oscillators is at equilibrium. This segment serves as a model of the reduced phase space in the case of incommensurate frequencies. Each physical observable is represented by some function $f(s_1)$ of a single variable, so the “effective dimension” of the reduced space is one.

The above Hamiltonian corresponds to a very degenerate integrable system. A more interesting scenario emerges upon deforming the constraint (\ref{H0}). In the action-angle variables above, consider, for example, the Hamiltonian
\begin{equation}\label{H}
H= \omega_1s_1+\omega_2s_2 +as_1^2+bs_2^2+2cs_1s_2\,.
\end{equation}
For sufficiently small values of the parameters $a$, $b$, and $c$, the isoenergetic surface $H=E>0$ of this Hamiltonian is still diffeomorphic to $S^3$. The functions (\ref{H12}) remain integrals of motion and foliate the isoenergetic surface into Liouville tori. The corresponding Hamiltonian vector field has the form
\begin{equation}\label{X}
X_{H}=\Omega_1(s_1,s_2)\frac{\partial}{\partial \phi_1}+\Omega_2(s_1,s_2)\frac{\partial}{\partial \phi_2}\,,
\end{equation}
where
\begin{equation}
\Omega_1=\omega_1+2as_1+2cs_2\,,\qquad \Omega_2=\omega_2+2bs_2+2cs_1\,.
\end{equation}
Since the oscillation frequencies become functions of the action variables, rational and irrational tori alternate with each other, forming separately dense sets.  As a result, all smooth functions on the isoenergetic surface preserved by the flow $X_{H}$ turn out to be constant on each torus, and each physical observable is represented by a function of a single variable, regardless of the algebraic properties of the numbers $\omega_1$, $\omega_2$, $a$, $b$, and $c$. Thus, as in the previously considered case of incommensurate frequencies, the effective dimension of the reduced phase space is one.

This last situation is, in some sense, generic. According to the Kolmogorov--Arnold--Moser theory, under a small perturbation of a non-degenerate integrable system, the corresponding Liouville tori do not disappear but merely deform, with non-resonant tori forming a dense set of positive measure \cite{Arn}. From the perspective of Hamiltonian systems with constraints, this result can be interpreted as follows. If a constraint $T\approx 0$ defines the isoenergetic surface of a weakly perturbed integrable system in a $2n$-dimensional phase space, then the effective dimension of the reduced phase space is $n-1$.
In the opposite extreme, a Hamiltonian system may turn out to be ergodic on a certain isoenergetic surface. In this case, the effective dimension of the reduced phase space is zero, and the algebra of physical observables 
\(C^{\infty}(N)\) consists solely of constants.

\section{Quantum Reduction and Anomalies}

A universal tool for quantizing Hamiltonian systems with constraints is the Hamiltonian version of BRST theory\footnote{An acronym of the surnames Becchi, Rouet, Stora, and Tyutin.}, also known as the Batalin–Fradkin–Vilkovisky method \cite{HT, JS}.

Within the framework of the BRST-BFV formalism, the Poisson algebra of physical observables $C^\infty(N)$ is realized as the cohomology algebra of a certain differential graded algebra ${A}=\bigoplus_{n\in\mathbb{Z}}A_n$ with differential $\delta: A_n\rightarrow A_{n+1}$. More specifically: the original phase space of the system $M$ is identified with the base of a $\mathbb{Z}$-graded manifold $\mathcal{M}$, whose structure sheaf $\mathcal{O}_{\mathcal{M}}$ is generated by two sets of generators $\mathcal{C}^a$ and $\mathcal{P}_a$ of degree $1$ and $-1$, respectively. These generators can conveniently be thought of as odd coordinates on $\mathcal{M}$. The original Poisson bracket extends from $M$ to $\mathcal{M}$ canonically:
\begin{equation}
\{\mathcal{C}^a,\mathcal{C}^b\}=0\,,\qquad \{\mathcal{C}^a,\mathcal{P}_b\}=\delta_b^a\,,\qquad \{\mathcal{P}_a,\mathcal{P}_b\}=0\,.
\end{equation}
In the Poisson algebra of smooth functions $C^\infty(\mathcal{M})$, one introduces a degree-one function:
\begin{equation}\label{Q}
Q = \mathcal{C}^aT_a + \frac12\mathcal{P}_a U^a_{bc}\mathcal{C}^b \mathcal{C}^c + \mathcal{O}(\mathcal{C}^3)\,,
\end{equation}
which satisfies the \textit{classical master equation}:
\begin{equation}\label{meq}
\{Q, Q\}=0\,.
\end{equation}
The function $Q$ is determined by the constraints $T_a\approx 0$ uniquely up to a canonical transformation on $\mathcal{M}$ and is called the {\it BRST charge}. In particular, isolating in (\ref{meq}) the terms linear in $\mathcal{C}^a$, one finds:
\begin{equation}
\{T_a,T_b\}=U_{ab}^c T_c\,,
\end{equation}
which precisely reproduces the involutivity condition  (\ref{inv}).

It also follows from equation (\ref{meq}) that the Hamiltonian vector field $\delta=X_Q$ satisfies the nilpotency condition $\delta^2=0$. Since the operator $\delta: C^{\infty}(\mathcal{M})\rightarrow C^{\infty}(\mathcal{M})$ has degree one and differentiates both the pointwise multiplication of functions and the Poisson bracket, one can speak of the differential graded Poisson algebra ${A}=(C^{\infty}(\mathcal{M}), \{-,-\},\delta)$. The corresponding cohomology groups $H({A})=\bigoplus_{n\in \mathbb{Z}} H^n({A})$ naturally inherit the structure of a graded Poisson algebra and play a central role in BRST theory. In particular, the Poisson subalgebra $H^0(A)$ is identified with the algebra of physical observables.

To establish the equivalence of this definition with the one from the previous section, it is enough to note that each $\delta$-cocycle $F$ of degree zero is given by a function
\begin{equation}\label{F}
F=f+\mathcal{P}_aV^a_b\mathcal{C}^b+\mathcal{O}(\mathcal{C}^2)\,,
\end{equation}
where $f$, $V^a_b$ are smooth functions on $M$. The closedness condition
\begin{equation}\label{dF}
\delta F=\{Q,F\}=0
\end{equation}
leads to a chain of relations on the structural functions, beginning with 
\begin{equation}
\{T_a,f\}=V_a^bT_b\,.
\end{equation}
This equation, being equivalent to the defining condition (\ref{CMJ}), allows one to interpret $f$ as an element of the subalgebra $C^\infty(M)^J$. The passage from cocycles to cohomology classes is equivalent to factoring $C^\infty(M)^J$ by the ideal $J$. Conversely, every element $f\in C^\infty(M)^J$ lifts to a certain $\delta$-cocycle (\ref{F}), and two $\delta$-cocycles $F$ and $F'$ are cohomologous if and only if $f-f'\in J$. This establishes the isomorphism $C^\infty(M)^J/J\simeq H^0(A)$ of Poisson algebras\footnote{Since the Hamiltonian of the system belongs to the physical observables, the condition of its BRST invariance $\{Q,H\}=0$ can be interpreted as the conservation of $Q$ over time. Hence the name -- “BRST charge.”}.

The BRST-BFV formalism described above proves especially useful in constructing the quantum theory of Hamiltonian systems with constraints. Below we consider the deformation version of quantization \cite{Fed}. Recall that in the deformation quantization approach, the set of observables is identified with the elements of the tensor product $\mathfrak{A}=C^\infty(\mathcal{M})\otimes \mathbb{C}[[\hbar]]$, where the right factor corresponds to the space of formal power series in $\hbar$ with complex coefficients. The space $\mathfrak{A}$ is endowed with an associative multiplication mimicking operator multiplication, the so-called $\ast$-product. The latter is a formal deformation of the commutative multiplication in the algebra of functions $C^\infty(\mathcal{M})$ and has the following form:
\begin{equation}
F\ast G=F\cdot G+\sum_{k=1}^\infty \hbar^kD_k(F,G)\,.
\end{equation}
Here $D_k(F,G)$ are certain bi-differential $\mathbb{C}[[\hbar] ]$-linear operators, with
\begin{equation}\label{D1}
D_1(F,G)-(-1)^{|F||G|}D_1(G,F)=i\{F,G\}
\end{equation}
($|F|$ denotes the degree of the function $F$ with respect to the $\mathbb{Z}$-grading). The associative graded algebra $(\mathfrak{A}, \ast)$ is called the algebra of {\it quantum observables}.

When moving to the quantum description,  the classical master equation (\ref{meq}) is replaced by its quantum counterpart
\begin{equation}\label{qmeq}
Q\ast Q=0\,,
\end{equation}
and the BRST invariance of a physical observable (\ref{dF}) takes the form
\begin{equation}\label{qclos}
[Q,F]_\ast:=Q\ast F-(-1)^{|Q||F|}F\ast Q=0\,.
\end{equation}
The adjoint action of $Q$ on the algebra of quantum observables turns $\mathfrak{A}$ into a differential graded algebra, allowing one to speak of the graded cohomology algebra $H(\mathfrak{A})=\bigoplus_{n\in\mathbb{Z}} H^n(\mathfrak{A})$. The subalgebra $H^0(\mathfrak{A})$ is identified with the algebra of physical observables of the quantum theory.

The correspondence principle between classical and quantum mechanics implies that the quantum BRST charge and physical observables are given by formal series:
\begin{equation}
Q=Q_0+\hbar Q_1+\hbar^2 Q_2+\ldots\,, \qquad F=F_0+\hbar F_1+\hbar^2F_2+\ldots \,,
\end{equation}
where $Q_0$ and $F_0$ now denote the classical BRST charge (\ref{Q}) and the classical observable (\ref{F}). The additional terms proportional to powers of $\hbar$ are treated as possible “quantum corrections” to classical quantities, vanishing in the classical limit $\hbar\rightarrow 0$. Expanding equations (\ref{qmeq}) and (\ref{qclos}) in power series in $\hbar$, we obtain two infinite chains of equations for the quantum corrections:
\begin{equation}\label{QF}
\delta Q_n=-\!\!\!\sum_{k+l+s=n+1} \!\!\!D_k(Q_{l},Q_s)\,, \qquad
\delta F_n=\!\!\!\sum_{{k+l+s=n+1}}\!\!\!D_k(F_l,Q_s)-D_k(Q_s,F_l)\,.
\end{equation}
Simple induction on $n$ shows that the sums on the right-hand sides of the equations are $\delta$-cocycles of degree 2 and 1, respectively, provided that all previous equations have been identically satisfied. The triviality of these cocycles is a necessary and sufficient condition for the solvability of the $n$-th equation in the chains (\ref{QF}). In particular, the triviality of the second cohomology group $H^2(A)$ guarantees the existence of a quantum BRST charge, and the vanishing of $H^1(A)$ ensures the existence of a quantum physical observable $F$ for any classical physical quantity $F_0$. In other words, the groups $H^2(A)$ and $H^1(A)$ contain all potential obstructions to solving the quantum master equation (\ref{qmeq}) and the quantum closedness condition (\ref{qclos}). In cases where these obstructions indeed materialize and quantum Hamiltonian reduction becomes impossible\footnote{Note that quantum reduction for second-class constraints does not encounter cohomological obstructions, at least within the deformation quantization approach \cite{BGL}.}, one speaks of “quantum anomalies.”

Below we will construct an explicit example of a Hamiltonian system with a single constraint and a nontrivial group $H^1(A)$. Moreover, we will present a concrete deformation quantization scheme realizing certain elements of $H^1(A)$ as insurmountable obstructions to solving the quantum closure condition (\ref{qclos}).

\begin{figure}
    \centering
\begin{tikzpicture}[scale=1.2]
\draw[red, -] (-1.9,-2.4) -- (1.1,2.4);
\draw[red, -] (-2.7,-0.9) -- (2.2,0.85);
\draw[->] (-0.7,-2.5) -- (-0.7,2.5);
\draw[->] (-2.9,-0.4) -- (2.5, -0.4);
\draw[blue] (-2.15,-1.4) rectangle (1.15,1.13);
\draw[thick, rotate around={30:(0,-1)}] (0, 0) ellipse (1.8 and 1);
\filldraw[black] (0.3,1.11) circle (1pt) node[above]{${\!\!}_{T_2}$};
\filldraw[black] (1.15,0.48) circle (1pt) node[above right ]{${}^{T_1}{}$};
\filldraw[black] (-0.7,0.9) circle (1pt);
\filldraw[black] (0.75,-0.4) circle (1pt);
\draw (-0.62,-0.3) node[below left]  {${}^{\small 0}$};
\draw (1.5,0.5) node [right] {${}_{\Omega_1=0}$};
\draw (0.8,2) node [right] {${}_{\Omega_2=0}$};
\draw (2.5, -0.4) node [below] {${}_{s_1}$};
\draw (-0.6, 2.3) node [left] {${}_{s_2}$};
\draw (-1, 0.8) node [left] {${}_{H=E}$};
\draw[very thick] (4.5,0)--(5,1)--(6,-0.4);
\filldraw[black]  (4.5,0) circle (1.5pt);
\filldraw[black]  (6,-0.4) circle (1.5pt);
\filldraw[fill=white, thick]  (5,1) circle (1.5pt);
\filldraw[black]  (-2.15,-0.7) circle (1pt);
\filldraw[black]  (-1.27,-1.4) circle (1pt);
\draw (5.8, -1) node [left] {${\Gamma(s_2, \Sigma)}$};
\draw (5.1, 1.1) node [above] {${}_{T_2}$};
\end{tikzpicture}
\caption{Left: projection of the isoenergetic surface $\Sigma$ of the Hamiltonian (\ref{H}) onto the plane of action variables. Right: the graph  depicting the bifurcation of the integral $s_2$ on the isoenergetic surface $\Sigma$.}
    \label{P1}
\end{figure}
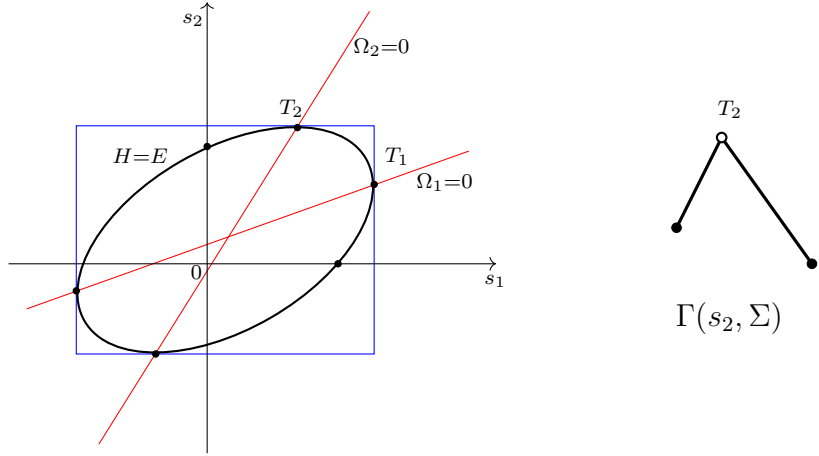

\vspace{3mm}
\paragraph{\bf Example.} Consider the 4-dimensional phase space $\mathbb{R}^4$ with the canonical Poisson bracket and the first-class constraint $T=H-E\approx 0$, where the function $H$ is given by expression (\ref{H}). Thus, the algebra of functions on the extended phase space $\mathcal{M}$ is generated by the canonical variables $(x_1, p_1; x_2, p_2; \mathcal{C}, \mathcal{P})$. Introducing a unified notation for the canonical coordinates on $\mathbb{R}^4$,
$$(\xi^1,\xi^2, \xi^3,\xi^4)=(x_1, p_1, x_2, p_2)\,,$$ 
we define the following $\ast$-product:
\begin{equation}\label{FG}
F\ast G=F,\exp\left(\frac{\stackrel{\leftarrow}{\partial}}{\partial \xi^A}\hbar W^{AB}\frac{\stackrel{\rightarrow}{\partial}}{{\partial} \xi^B}+\frac{\stackrel{\leftarrow}{\partial}}{\partial \mathcal{C}}\hbar\frac{\stackrel{\rightarrow}{\partial}}{\partial \mathcal{P}} \right)G\,,
\end{equation}
where
\begin{equation}
W=\frac12\left(
\begin{array}{cccc}
\alpha& i &0 &0\\
-i& \alpha^{-1}&0&0\\
0&0&\beta&i\\
0&0&-i&\beta^{-1}
\end{array}\right)\,,
\end{equation}
and $\alpha$ and $\beta$ are some nonzero real constants. The arrows above the derivatives show which functions the derivatives act on. The derivative with respect to the odd variable $\mathcal{C}$ is right, while that with respect to $\mathcal{P}$ is left. The form $W$ is clearly Hermitian and satisfies condition (\ref{D1}). The rank of the form $W$ is two. Its left kernel is spanned by the differentials
\begin{equation}
\lambda_1=i dx_1-\alpha dp_1\,,\qquad \lambda_2=i dx_2-\beta dp_2\,.
\end{equation}
The covectors $\lambda_{1,2}$ and their complex conjugates $\bar \lambda_{1,2}$ define a decomposition of the complexified phase space  into a direct sum of Lagrangian subspaces $L\oplus \bar L$. When restricted to the subalgebra $C^\infty(\mathbb{R}^4)\otimes \mathbb{C}[[\hbar]]$, the product (\ref{FG}) defines a {\it Wick-type} deformation quantization of the original phase space\footnote{The Wick $\ast$-product is a geometric version of creation-annihilation operators and is often used in quantizing oscillator systems. For more on Wick quantization, see \cite{BS, DLS, KS}.}. Note that the Hamiltonian flow $X_H$ does not preserve the Wick polarization $L\oplus \bar L$, while the flow $X_{H_0}$ corresponding to the harmonic oscillator (\ref{H0}) preserves this polarization only in the special case $\alpha=\beta=1$.

For a single first-class constraint, the classical BRST charge automatically satisfies the quantum master equation without any quantum corrections:
\begin{equation}
Q=\mathcal{C}(H-E)\,,\qquad Q\ast Q =0\,,
\end{equation}
and we adopt this expression as the quantum BRST charge. The general expression for a physical observable $F$ takes the following form:
\begin{equation}
F=f+\hbar f_1+\mathcal{P}(V+\hbar V_1)\mathcal{C}+\mathcal{O}(\hbar^2)\,.
\end{equation}
Substituting this into (\ref{QF}), we obtain a chain of equations, the first two of which are given by
\begin{equation}
X_H f=(H-E)V\,,
\end{equation}
\begin{equation}\label{f1}
X_H f_1=D_2(H,f)-D_2(f,H)+D_1(H,V)+(H-E)V_1\,.
\end{equation}
The first equation characterizes $f$ as a classical physical observable. The second equation determines the first quantum correction to $f$. Below we will show that, in the general case, this equation for $f_1$ has no solution in the class of smooth functions and, consequently, the algebra of physical observables in the quantum theory is strictly smaller than in the classical theory.
Let us take the function 
\begin{equation}\label{f}
f=\frac12(p_1^2+x_1^2)=s_1
\end{equation}
as a classical physical observable. Then $V=0$ and equation (\ref{f1}) simplifies accordingly. Substituting expressions (\ref{H}) and (\ref{f}), we find
\begin{equation}\label{Xf1}
X_H f_1\approx - 2i \, a (\alpha-\alpha^{-1})x_1p_1\,.
\end{equation}
In the general case, this equation has no smooth solutions. We assume that $\alpha\neq 1$ and the matrix
\begin{equation}
\left(\begin{array}{cc}
a & c \\
c & b
\end{array}\right)\,,
\end{equation}
defining the quadratic form in (\ref{H}), is positive definite. This means that the constraint equation $H(s_1,s_2)=E$ defines an ellipse in the plane of variables $s_1$ and $s_2$, see Fig. \ref{P1}. We interpret the constraint surface as an isoenergetic surface of the Hamiltonian $H$. The level surfaces of the integral $s_2$ define a foliation of the isoenergetic surface into Liouville tori. The corresponding Fomenko graph \cite{Fom} is shown in Fig. \ref{P1} on the right. The black vertices correspond to the points $s_1=0$ and $s_2=0$, where the tori degenerate into circles. The white vertex corresponds to the maximum of the function $s_2$ on the isoenergetic surface (point $T_2$). At this point, two Liouville tori merge into one (maximal) torus. Point $T_1$ corresponds to the merging of two Liouville tori for the integral $s_1$. The main observation is that both maximal tori are resonant. Indeed, at point $T_2$ of the isoenergetic surface we have
\begin{equation}
\frac{\partial s_2}{\partial s_1}= \frac{\partial H/\partial s_1}{\partial H/\partial s_2}=\frac{\Omega_1}{\Omega_2}=0\qquad \Rightarrow\qquad \Omega_1=0\,,\qquad \Omega_2\neq 0\,.
\end{equation}
Similarly, at point $T_1$ we have $\Omega_2=0$ and $\Omega_1\neq 0$. In action-angle variables, the Hamiltonian vector field (\ref{X}) on the maximal torus $T_2$ takes the form $X_H=\Omega_2\partial/\partial\varphi_2$. All integral trajectories of this vector field are obviously circles. Restricting equation (\ref{Xf1}) to the maximal torus $T_2$ and introducing the notation $\tilde f_1=f_1|_{T_2}$, we obtain
\begin{equation}
\Omega_2\frac{\partial \tilde{f}_1}{\partial\varphi_2} = - 2i \, a (\alpha-\alpha^{-1})s_1 \sin 2\varphi_1\,.
\end{equation}
Integrating this equation over $\varphi_2$ from $0$ to $2\pi$, we arrive at a contradiction.

The geometric idea behind the above proof  is that we have found a closed integral curve $\gamma$ of the vector field $X_H$, lying on the constraint surface, such that 
\begin{equation}
\Delta=\left.\big[D_2(H,f)-D_2(f,H)+D_1(H,V)\big]\right|_{\gamma}\neq 0\,. \end{equation}
Restricting equation (\ref{f1}) to this trajectory gives $X_H(f_1|_\gamma)=\Delta$. This equality is contradictory because any smooth function $f_1|_\gamma$ on a circle has extremum points where $X_H(f_1|_\gamma)=0$.

It also follows from this reasoning that if the vector field $X_H$ has closed integral trajectories, then the corresponding Hamiltonian system with the constraint $H-E\approx 0$ has a nonzero cohomology group $H^1(A)$. For example, a nontrivial 1-cocycle is given by the function $\mathcal{C}$. According to the Weinstein conjecture, which is now a theorem, every compact energy hypersurface $\Sigma\subset \mathbb{R}^{2n}$ of contact type has at least one closed trajectory (see \cite{P}, \cite{V}).  

\section{Discussion}

In this note, we have constructed a simple example of a finite-dimensional Hamiltonian system with constraints that exhibit unavoidable quantum anomalies. These anomalies cause the algebra of quantum physical observables to be strictly smaller than that of classical observables.

The reader may wonder why we use the Wick-type deformation quantization scheme instead of the more traditional Weyl quantization. The reason is purely technical: for the Weyl $\ast$-product, the first quantum corrections to physical observables can always be chosen to vanish. Thus, the appearance of quantum anomalies is shifted to the next order in $\hbar$. The equation for the second quantum correction $f_2$ becomes more complicated and involves third derivatives of the functions $H$ and $f$. Quantum anomalies can arise only if $H$ and $f$ are Poisson-commuting polynomials of degree greater than two in the canonical variables.

In our example, the action variables $s_1$ and $s_2$ play fully symmetric roles. Therefore, if both points $T_1$ and $T_2$ in Fig. \ref{P1} lie in the first quadrant of the coordinate plane and the parameters $\alpha$ and $\beta$ are not equal to one, then the physical observables $s_1$ and $s_2$ cannot both be quantized simultaneously. At the same time, for $\alpha \neq 1$ and $\beta = 1$, the quantity $s_2$ can be quantized, while $s_1$ cannot. This may seem strange, since the effective dimension of the reduced phase space is one, meaning that any physical quantity should be representable as a function of a single variable. However, it should be noted that, unlike the degenerate Hamiltonian $H_0$, the variables $s_1$ and $s_2$ are not global coordinates on the reduced phase space of the system (which corresponds to the arc of the ellipse lying in the first quadrant of the $s_1 s_2$ coordinate plane). As a consequence, the action variables $s_1$ and $s_2$ are not well-defined functions of each other on the constraint surface. The presence of the maximal tori $T_1$ and $T_2$ precisely obstructs this. As a result, $s_1$ and $s_2$ become distinct physical observables that are functionally independent of each other. This explains why one of these physical quantities can be quantized while the other cannot.

In conclusion, we note that the problem of quantum Hamiltonian reduction by a single first-class constraint is closely related to the quantization of contact manifolds. Indeed, if the original symplectic manifold $M$ was exact (i.e., admitted a presymplectic potential $\theta$), then under certain regularity conditions, the restriction of the 1-form $\theta$ to the constraint surface $\Sigma \subset M$ endows the latter with the structure of a contact manifold. Deformation quantization of contact and contact-metric manifolds has been developed in recent works \cite{ES1, ES2}, where certain obstructions to quantizing physical quantities in the presence of closed characteristics have also been identified. From this perspective, the energy surface $H = E$ with contact form $\theta = p_1 dx_1 + p_2 dx_2$ provides an example of a contact-metric manifold whose Wick-type deformation quantization leads to anomalies.


\begin{thebibliography}{10}


\bibitem{Dirac} 
P. A. M. Dirac, \textit{Lectures on Quantum Mechanics}, Belfer Graduate School Monograph Series 2, 1964.

\bibitem{BF} I. A. Batalin, E. S. Fradkin, \textit{Operatorial quantization of dynamical systems subject to constraints. A further study of the construction}, Annales de l'I.H.P. Physique théorique, Tome 49, no. 2, pp. 145-214, 1988.

\bibitem{HT} M. Henneaux, C. Teitelboim, \textit{Quantization of Gauge Systems}, Princeton University Press, 1992, 520 p. 


\bibitem{BHW} M. Bordemann, HC. Herbig, and S. Waldmann, \textit{BRST cohomology and phase space reduction in deformation quantization},  Communications in Mathematical Physics, 210, pp. 107-144, 2000.

\bibitem{LS} S. L. Lyakhovich, A. A.  Sharapov, \textit{Characteristic classes of gauge systems}, Nuclear Physics B, 703(3), pp. 419-453, 2004.

\bibitem{LMS} S. L. Lyakhovich, E. A. Mosman, A. A. Sharapov, \textit{Characteristic classes of Q-manifolds: Classification and applications}, Journal of Geometry and Physics, Volume 60, Issue 5, pp. 729-759, 2010. 


\bibitem{Fom} A. T. Fomenko, \textit{Topological invariants of Liouville integrable Hamiltonian systems},   Funct. Anal. Appl., 22:4, pp. 286–296, 1988.

\bibitem{Arn}V. I. Arnol'd, \textit{Small denominators and problems of stability of motion in classical and celestial mechanics},  Russian Math. Surveys, 18:6, pp. 85–191, 1963.


\bibitem{JS} J. Stasheff, \textit{Homological reduction of constrained Poisson algebras}, Journal of Differential Geometry, 45.1, pp. 221-240, 1997.

\bibitem{Fed} B. V.~Fedosov, \textit{Deformation Quantization and Index Theory}, Akademie-Verlag, Berlin, Germany, 325 p, 1996. (Mathematical topics: 9).


\bibitem{BGL}I. A. Batalin, M. A Grigoriev, and S. L. Lyakhovich, \textit{Star Product for Second-Class Constraint Systems from a BRST Theory},  Theoret. and Math. Phys., 128:3, pp. 1109–1139, 2001.


\bibitem{BS}  F. A. Berezin, M. A. Shubin,
{\it The Schr\"odinger Equation}, Springer Science \& Business Media, 555 p, 1991 (Mathematics and its Applications, 66). 


\bibitem{DLS} V. A. Dolgushev, S. L. Lyakhovich, A. A. Sharapov, \textit{Wick type deformation quantization of Fedosov manifolds}, Nuclear Physics B, Volume 606, Issue 3, pp. 647-672, 2001.


\bibitem{KS} A. V. Karabegov, M. Schlichenmaier, \textit{Almost-Kähler Deformation Quantization}, Letters in Mathematical Physics, 57, pp. 135–148, 2001.

\bibitem{ES1} B. M. Elfimov, A. A. Sharapov, \textit{Deformation quantization of contact manifolds}, Letters in Mathematical Physics, 112.6, 124, 2022.   

\bibitem{ES2} B. M.~Elfimov, A. A. Sharapov, \textit{Wick-type deformation quantization of contact metric manifolds}, Letters in Mathematical Physics, 114.2, 37, 2024.

\bibitem{P} F.~Pasquotto,  \textit{A short history of the Weinstein conjecture}, Jahresbericht der Deutschen Mathematiker-Vereinigung, 114, 119-130 (2012).

\bibitem{V} C.~Viterbo, \textit{ A proof of Weinstein’s conjecture in $\mathbb{R}^{2n}$}, Ann. Inst. Henri Poincaré, Anal. Non Linéaire, 4, 337–356 (1987).



\end{thebibliography}
\end{document}